\documentclass[11pt, a4paper]{article}
\pdfoutput=1
\usepackage{jcappub}

\usepackage{booktabs}
\usepackage{multirow}
\usepackage{amssymb}
\usepackage[export]{adjustbox}
\usepackage{floatrow}
\newfloatcommand{capbtabbox}{table}[][3.5in]
\newfloatcommand{capbfigbox}{figure}[][2.3in]
\usepackage{blindtext}
\usepackage{amsmath}
\usepackage{booktabs}
\usepackage{aas_macros}

\def\lsim{\mathrel{\raise.3ex\hbox{$<$\kern-.75em\lower1ex\hbox{$\sim$}}}}
\def\gsim{\mathrel{\raise.3ex\hbox{$>$\kern-.75em\lower1ex\hbox{$\sim$}}}}

\begin{document}

\hspace*{110mm}{\large \tt FERMILAB-PUB-17-530-A}

\vskip 0.2in

\title{Measuring the Local Diffusion Coefficient with H.E.S.S. Observations of Very High-Energy Electrons} 





\author{Dan Hooper$^{a,b,c}$}\note{ORCID: http://orcid.org/0000-0001-8837-4127}
\emailAdd{dhooper@fnal.gov}
\author{and Tim Linden$^d$}\note{ORCID: http://orcid.org/0000-0001-9888-0971}
\emailAdd{linden.70@osu.edu}

\affiliation[a]{Fermi National Accelerator Laboratory, Center for Particle Astrophysics, Batavia, IL 60510}
\affiliation[b]{University of Chicago, Department of Astronomy and Astrophysics, Chicago, IL 60637}
\affiliation[c]{University of Chicago, Kavli Institute for Cosmological Physics, Chicago, IL 60637}
\affiliation[e]{Ohio State University, Center for Cosmology and AstroParticle Physics (CCAPP), Columbus, OH  43210}

\abstract{The HAWC Collaboration has recently reported the detection of bright and spatially extended multi-TeV gamma-ray emission from Geminga, Monogem, and a handful of other nearby, middle-aged pulsars. The angular profile of the emission observed from these pulsars is surprising, in that it implies that cosmic-ray diffusion is significantly inhibited within $\sim$25~pc of these objects, compared to the expectations of standard Galactic diffusion models. This raises the important question of whether the diffusion coefficient in the local interstellar medium is also low, or whether it is instead better fit by the mean Galactic value. Here, we utilize recent observations of the cosmic-ray electron spectrum (extending up to $\sim$20~TeV) by the H.E.S.S. Collaboration to show that the local diffusion coefficient cannot be as low as it is in the regions surrounding Geminga and Monogem. Instead, we conclude that cosmic rays efficiently diffuse through the bulk of the local interstellar medium. Among other implications, this further supports the conclusion that pulsars significantly contribute to the observed positron excess.}

\maketitle

\section{Introduction}
\label{sec:introduction}

Measurements of the cosmic-ray positron fraction by the PAMELA~\cite{Adriani:2010rc} and AMS-02~\cite{Aguilar:2013qda} experiments (as well as by HEAT~\cite{Barwick:1997ig}, AMS-01~\cite{Aguilar:2007yf} and Fermi~\cite{FermiLAT:2011ab}) have identified an excess relative to the standard predictions for secondary production in the interstellar medium (ISM). This indicates that significant quantities of $\sim$0.01-1 TeV positrons must be produced as primary cosmic rays. Because high-energy positrons efficiently cool through a combination of synchrotron and inverse-Compton processes, at least some of these positrons must be produced relatively nearby. Proposals for the origin of these particles include nearby pulsars~\cite{Hooper:2008kg, Yuksel:2008rf, Profumo:2008ms, Malyshev:2009tw, Grasso:2009ma, Linden:2013mqa,Cholis:2013psa}, annihilating dark matter~\cite{Bergstrom:2008gr, Cirelli:2008jk, Cholis:2008hb,  Cirelli:2008pk, Nelson:2008hj, ArkaniHamed:2008qn, Cholis:2008qq, Cholis:2008wq, Harnik:2008uu, Fox:2008kb, Pospelov:2008jd, MarchRussell:2008tu, Chang:2011xn,Cholis:2013psa}, and the acceleration of secondary positrons in nearby supernova remnants~\cite{Blasi:2009hv, Mertsch:2009ph, Ahlers:2009ae,Cholis:2013lwa,Kachelriess:2011qv, Kachelriess:2012ag,Cholis:2017qlb}. 

Earlier this year, the scientific collaboration operating the High-Altitude Water Cherenkov (HAWC) Observatory released their first measurements of the very-high energy gamma-ray emission from the nearby pulsars Geminga and Monogem~\cite{Abeysekara:2017hyn} (see also Refs.~\cite{2015arXiv150803497B,2016JPhCS.761a2034C,2015arXiv150907851P}). Even more recently, HAWC has reported that the emission from these sources follows a diffusive profile extending out to at least $\sim$5$^\circ$ in radius (corresponding to a physical extent of $\sim$25~pc)~\cite{newhawc}. The spatially extended nature of this emission indicates that it is generated through the inverse Compton scattering of very high-energy (VHE) electrons\footnote{Throughout this paper, unless otherwise stated, we refer to both electrons and positrons as simply ``electrons''.} with the cosmic microwave background and other radiation fields. Among other reasons, this result is important because it suggests that HAWC will likely be able to detect the ``TeV Halos'' around many pulsars, including those whose radio and GeV emission are not beamed in our direction~\cite{Linden:2017vvb}. Furthermore, the collective emission from the TeV Halos associated with the Milky Way's pulsar population is likely to generate much of the VHE gamma-ray emission observed from the Galactic Center~\cite{Hooper:2017rzt}, as well as the diffuse TeV excess previously reported by Milagro~\cite{Linden:2017blp}. 

Interestingly, the fluxes of VHE gamma-rays observed from Geminga and Monogem indicate that these sources inject a flux of positrons into the local ISM that is approximately equal to that required to account for the observed positron excess. Using the results from the HAWC Collaboration's 2HWC catalog~\cite{Abeysekara:2017hyn}, we argued previously that this information strongly favors the conclusion that the positron excess is generated by nearby pulsars, diminishing the motivation for annihilating dark matter or other exotic mechanisms~\cite{Hooper:2017gtd}.

A very different interpretation of this data has recently been put forth by the HAWC Collaboration~\cite{newhawc}. The angular profile of the VHE emission observed from Geminga and Monogem indicates that diffusion is extremely inefficient in the regions surrounding these sources, representing the first empirical determination of a diffusion coefficient in a $\sim$10-50 pc region within the local Galaxy. More quantitatively, they find that their data favors a diffusion coefficient (see Eq.~\ref{diffusionlosseq}) for the region surrounding Geminga that is $560^{+260}_{-170}$ times smaller than the value inferred from measurements of the boron-to-carbon ratio and other cosmic-ray secondary-to-primary ratios (which we take to be the GALPROP default value, $D\approx3.86 \times 10^{28} \, (E_e/{\rm GeV})^{0.33}$ cm$^2/$s~\cite{GALPROPSite}). Similarly, they find that the region surrounding Monogem requires a diffusion coefficient that is smaller than the standard value by a factor of $120^{+180}_{-90}$. The authors of Ref.~\cite{newhawc} assert that these reduced values for the diffusion coefficient are likely to be indicative of the diffusion coefficient throughout the bulk of the ISM, or at least within a sizable region surrounding the solar position. If true, the flux of cosmic-ray positrons that reaches the Solar System from these and other pulsars would be highly suppressed, suggesting that another -- and perhaps more exotic -- explanation would be required in order to explain the observed positron excess.

In this paper, we revisit this question and argue that the interpretation put forth in Ref.~\cite{newhawc} is incompatible with other cosmic-ray measurements, and in particular with the spectrum of cosmic-ray electrons reported by several experiments, including HESS~\cite{Aharonian:2008aa,Aharonian:2009ah,Abdalla:2017brm,hesstalk}, MAGIC~\cite{BorlaTridon:2011dk}, AMS-02~\cite{Aguilar:2014fea}, VERITAS~\cite{Staszak:2015kza}, and Fermi~\cite{Abdollahi:2017nat}. Of particular interest for the question at hand are the most recent measurements of the cosmic-ray electron spectrum from the HESS Collaboration~\cite{hesstalk}, which extend up to energies as high as $\sim$20 TeV~\cite{hesstalk}.\footnote{See \url{https://indico.snu.ac.kr/indico/event/15/session/5/contribution/694/material/slides/0.pdf}} Because VHE cosmic-ray electrons cool extremely rapidly, they provide critical information regarding the diffusion coefficient in the local ISM. In particular, at 20~TeV, cosmic-ray electrons cool on a timescale of $\sim$$10^4 \,{\rm years}$. If we adopt a standard value for the diffusion coefficient that is compatible with measurements of the boron-to-carbon and other secondary-to-primary ratios, we estimate that electrons will typically diffuse a distance of $\sim$$\sqrt{D t} \sim 200$ pc within this time. However, for the significantly smaller diffusion coefficient advocated in Ref.~\cite{newhawc}, the horizon for such VHE electrons is reduced to only $\sim$10--20 pc. As there are no plausible sources of VHE cosmic rays within this radius, we are forced to conclude that diffusion must be reasonably efficient throughout the majority of the local ISM, and that the conditions found in the regions surrounding Geminga and Monogem cannot be representative of the overall local Galactic environment.


\section{The Cosmic-Ray Electron Spectrum}

The diffusion and energy losses of cosmic-ray electrons can be described by the standard transport equation:
\begin{eqnarray}
\frac{\partial{}}{\partial{t}}\frac{dn_e}{dE_e}(E_e,\vec{x},t) =  \vec{\bigtriangledown} \cdot \bigg[D(E_e,\vec{x}) \vec{\bigtriangledown} \frac{dn_e}{dE_e}(E_e,\vec{x},t) \bigg] + \frac{\partial}{\partial E_e} \bigg[\frac{dE_e}{dt}(E_e) \, \frac{dn_e}{dE_e}(E_e,\vec{x},t)    \bigg]  + Q(E_e,\vec{x},t), \nonumber \\
\label{diffusionlosseq}
\end{eqnarray}
where $dn_e/dE_e$ is the differential number density of electrons, $D$ is the diffusion coefficient, and the source term, $Q$, describes the spectrum, distribution, and time profile of electrons injected into the ISM. Energy losses from inverse Compton and synchrotron processes are given by~\cite{Blumenthal:1970gc}:
\begin{eqnarray}
-\frac{dE_e}{dt}(r) &=& \sum_i \frac{4}{3}\sigma_T \rho_i(r) S_i(E_e) \bigg(\frac{E_e}{m_e}\bigg)^2 + \frac{4}{3}\sigma_T \rho_{\rm mag}(r) \bigg(\frac{E_e}{m_e}\bigg)^2  \\
& \approx & 1.02 \times 10^{-16} \, {\rm GeV}/{\rm s} \, \times \bigg[ \sum_i \frac{\rho_{i}(r)}{{\rm eV}/{\rm cm}^3} \, S_{i}(E_e) + 0.224 \,\bigg(\frac{B}{3\, \mu \rm{G}}\bigg)^2 \bigg]  \,  \bigg(\frac{E_e}{{\rm GeV}}\bigg)^2, \nonumber
%
%
\end{eqnarray}
where $\sigma_T$ is the Thomson cross section and the sum is carried out over the various components of the radiation backgrounds, consisting of the cosmic microwave background (CMB), infrared emission (IR), starlight (star), and ultraviolet emission (UV). Throughout our analysis, we adopt the following parameters: $\rho_{\rm CMB}=0.260$ eV/cm$^3$, $\rho_{\rm IR}=0.60$ eV/cm$^3$, \mbox{$\rho_{\rm star}=0.60$ eV/cm$^3$,} $\rho_{\rm UV}=0.10$ eV/cm$^3$, $\rho_{\rm mag}=0.224$ eV/cm$^3$ (corresponding to \mbox{$B=3\,\mu$G),} and $T_{\rm CMB} =2.7$ K, $T_{\rm IR} =20$ K, $T_{\rm star} =5000$ K and $T_{\rm UV} =$20,000 K. At very high energies ($E_e \gsim m^2_e/2T$), inverse Compton scattering is suppressed by the following Klein-Nishina factor~\cite{Longair}:
\begin{equation}
S_i (E_e) \approx \frac{45 \, m^2_e/64 \pi^2 T^2_i}{(45 \, m^2_e/64 \pi^2 T^2_i)+(E^2_e/m^2_e)}.
\end{equation}

The source term in Eq.~\ref{diffusionlosseq} includes contributions from individual sources of cosmic-ray electrons (pulsars, supernova remnants, etc.), as well as from the production of secondary particles. Secondary electrons and positrons are generated in the decays of pions and kaons that are produced in the collisions of hadronic cosmic rays with gas. The flux of cosmic-ray secondaries can be calculated from Eq.~\ref{diffusionlosseq} by setting $Q=\int J_p n_{\rm gas}(d\sigma/dE) dE_p$, where $J_p$ is the flux of hadronic cosmic rays, $n_{\rm gas}$ the gas density, and $d\sigma/dE$ is the differential cross section for the production of electrons and positrons~\cite{Moskalenko:1997gh}.

At the highest measured energies, cosmic-ray electrons cool on a timescale of $\sim$$3\times 10^4 \,{\rm years} \times (10\,{\rm TeV}/E_e)$, during which they diffuse a distance of only $\sim\sqrt{D t} \sim 300$ pc (for the diffusion coefficient inferred from measurements of the boron-to-carbon and other cosmic-ray secondary-to-primary ratios, $D \approx 3.86 \times 10^{28} \, (E_e/{\rm GeV})^{0.33}$ cm$^2/$s). In light of this, only a small volume contributes to the local VHE electron spectrum. Over this volume, the densities of both hadronic cosmic rays and gas are reasonably well known, allowing us to fairly reliably calculate the flux of VHE secondaries. Reducing the value of the diffusion coefficient will not substantially impact the local flux of VHE secondaries, because the cosmic-ray proton density is roughly homogeneous within this region.

\begin{figure}
\includegraphics[width=3.80in,angle=0]{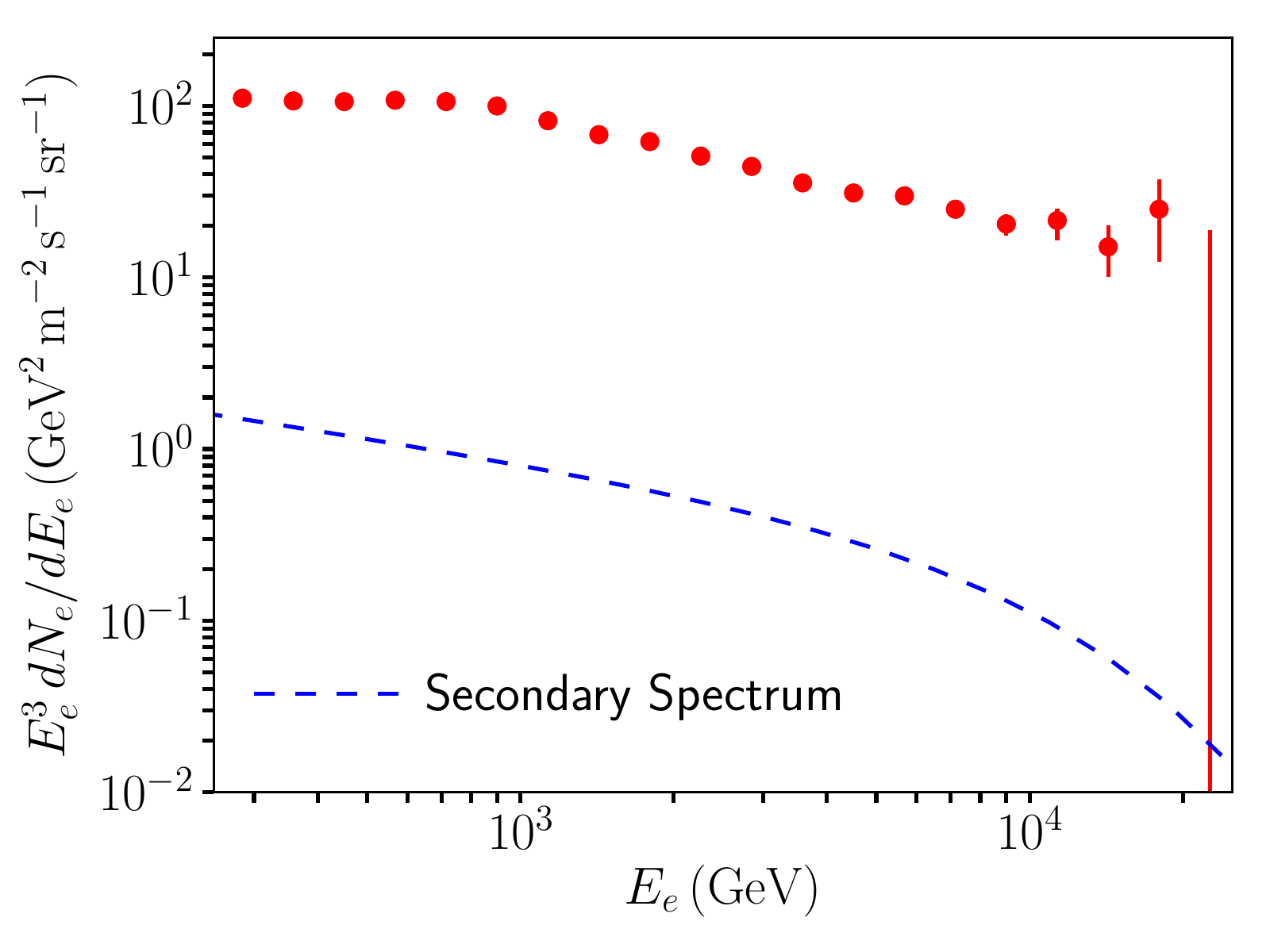}
\caption{The secondary contribution to the cosmic-ray electron (plus positron) spectrum, as calculated using GALPROP, and adopting the default values for all propagation parameters. Comparing this to the spectrum as reported by the HESS Collaboration, it is clear that secondary production provides only a small fraction of this flux. At the highest measured energies, this result is largely independent of the value of the diffusion coefficient.}
\label{secondary}
\end{figure}

In Fig.~\ref{secondary} we plot the spectrum of secondary electrons as predicted using the publicly available code GALPROP~\cite{GALPROPSite}. Here we have adopted the default parameter values, including a diffusion coefficient of $D=3.86 \times 10^{28} \, (E_e/{\rm GeV})^{0.33}$ cm$^2/$s. From this figure, it is clear that flux of  secondaries is quite small, making up only $\sim$1\% of the measured cosmic-ray electron spectrum. We thus conclude that secondaries contribute negligibly to the measured electron spectrum, especially at the highest measured energies. Moreover, we note that variations in the gas density, most importantly the local bubble, are likely to suppress the flux of cosmic-ray electrons even further, compared to the axisymmetric gas densities assumed in this model. This result, by itself, provides the first indication that a significant flux of $\sim$20~TeV primary electrons are produced in close proximity to the Solar System.


\begin{figure}
\includegraphics[width=3.80in,angle=0]{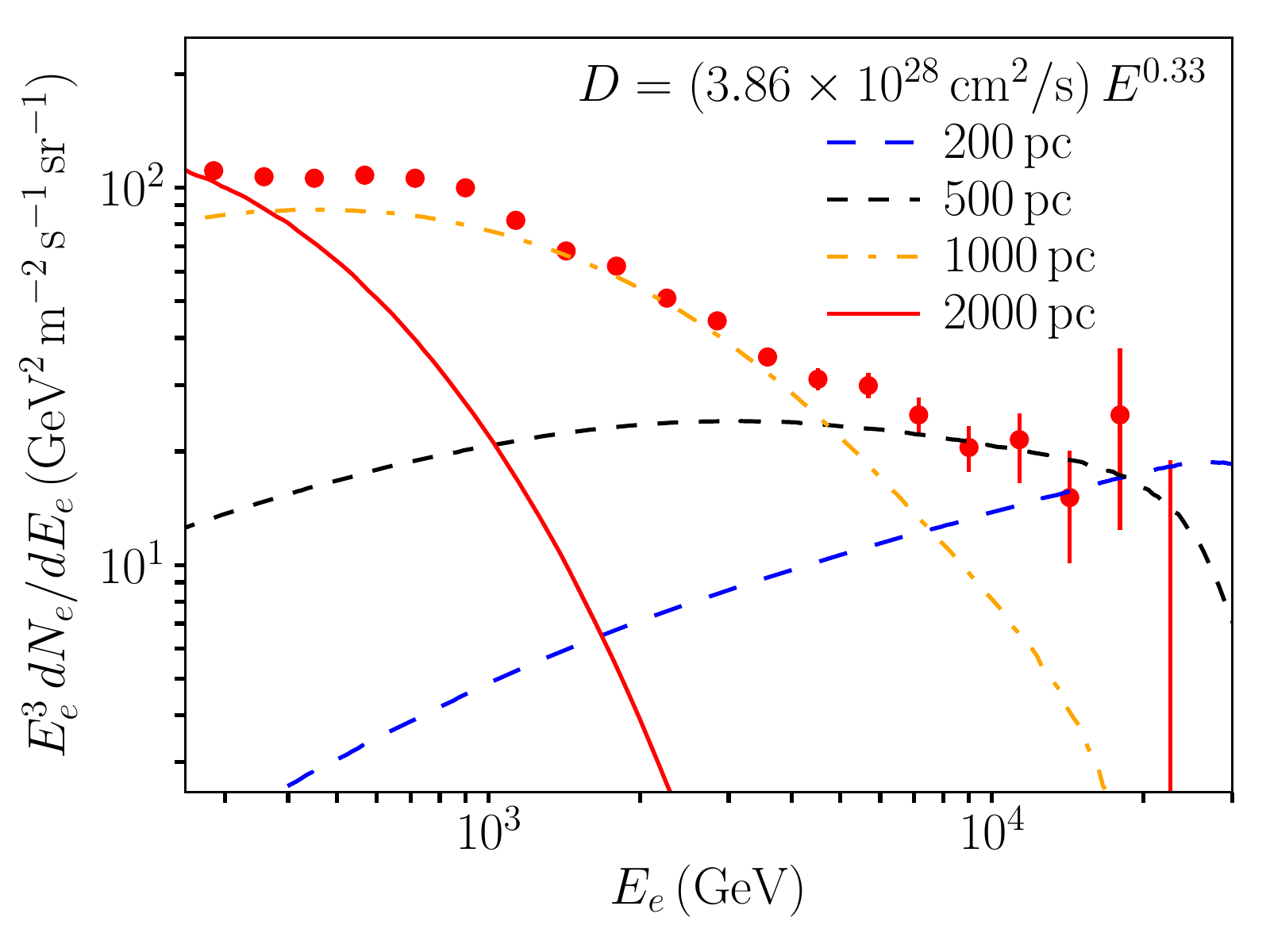}
\includegraphics[width=3.80in,angle=0]{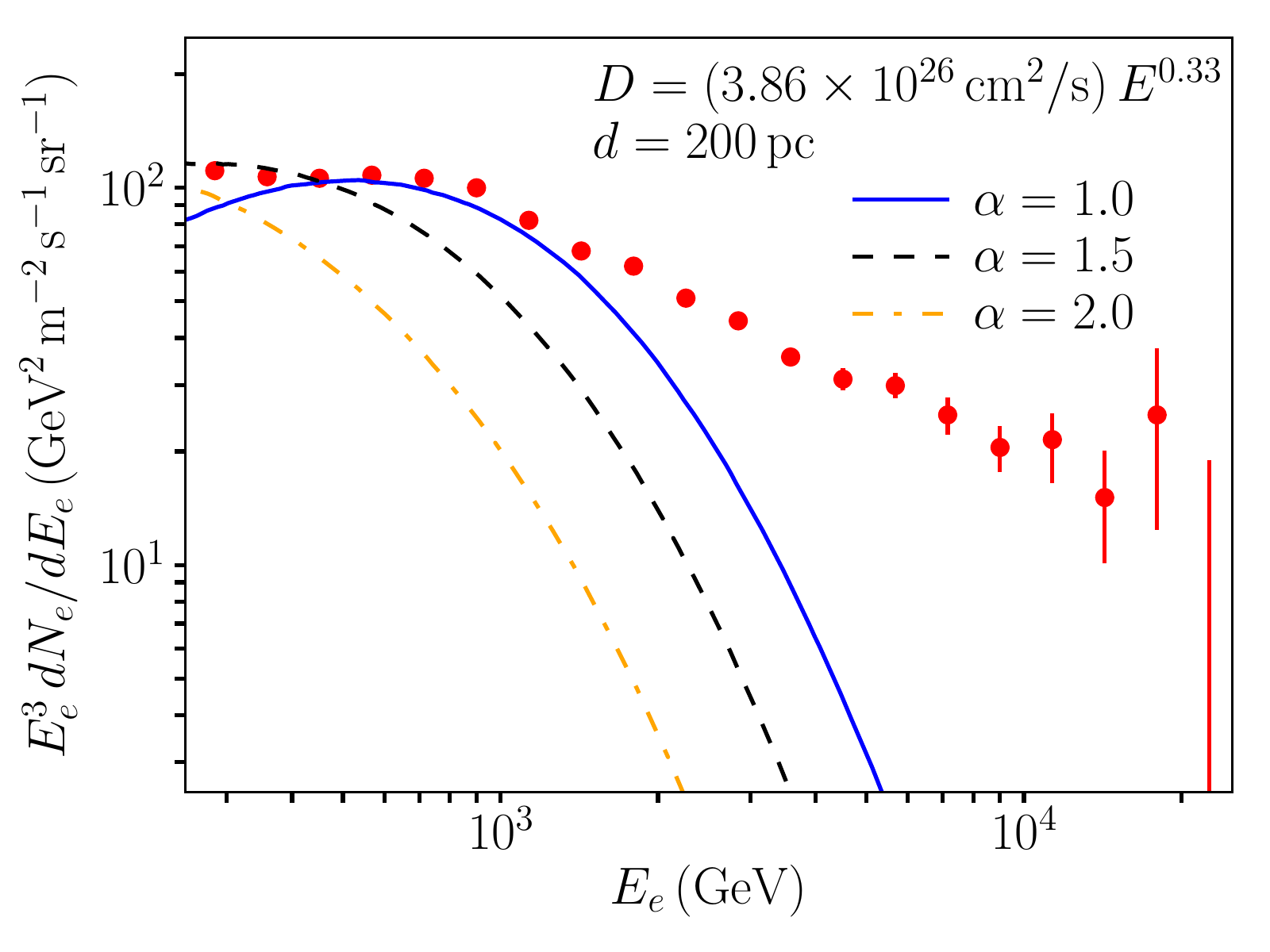}
\caption{The contribution to the cosmic-ray electron (plus positron) spectrum from individual nearby sources. In the top frame we plot the spectrum (arbitrarily normalized) of cosmic-ray electrons from a source located at various distances for a standard diffusion coefficient of $D=3.86 \times 10^{28} \, (E_e/{\rm GeV})^{0.33}$ cm$^2/$s and an injected spectral index of $\alpha=2$. In the lower frame, we plot the spectrum from a source 200 pc from the Solar System with a much lower diffusion coefficient in the range favored by HAWC for the regions surrounding the Geminga and Monogem, $D=3.86 \times 10^{26} \, (E_e/{\rm GeV})^{0.33}$ cm$^2/$s. If the ISM were described by such a low diffusion coefficient, this figure demonstrates that even very nearby sources could not account for the electron spectrum observed above $\sim$1-2 TeV. This is our main argument for why diffusion throughout the bulk of the ISM cannot be as inefficient as has been observed by HAWC in the regions surrounding Geminga and Monogem.}
\label{nearby}
\end{figure}

Next, we calculate the contribution to the cosmic-ray electron spectrum from individual nearby sources. In the top frame of Fig.~\ref{nearby}, we plot the spectral shape (arbitrarily normalized) of cosmic-ray electrons from a source located at distances between 200 and 1000 pc, for a standard diffusion coefficient of $D=3.86 \times 10^{28} \, (E_e/{\rm GeV})^{0.33}$ cm$^2/$s. For the spectrum that is injected from these sources, we have adopted a power-law form, $dN_e/dE_e \propto E^{-\alpha}$, with an index of $\alpha=2$. It is clear from this figure that the observed spectrum requires the presence of sources within approximately $\sim$$500$ pc of the Solar System, at least for this choice of the diffusion coefficient and spectral index.

In the lower frame of Fig.~\ref{nearby}, we show the primary electron spectrum generated by a source that is located only 200~pc from the Solar System, assuming a significantly lower diffusion coefficient of $D=3.86 \times 10^{26} \, (E_e/{\rm GeV})^{0.33}$ cm$^2/$s, as advocated by the HAWC Collaboration.\footnote{Throughout, we have adopted a maximum injected energy of 1 PeV. Although most of our results are not sensitive to this choice, the case in which $\alpha=1$ (in the lower frame of Fig.~\ref{nearby}) is an exception.} These results indicate that, if the local ISM were described by such a low diffusion coefficient, the nearest known high-energy sources would be unable to explain the HESS electron flux above $\sim$1-2~TeV, even if the source spectrum were as hard as $\alpha$~=~1. This is our primary argument explaining why diffusion throughout the bulk of the local ISM cannot be as inefficient as implied by HAWC's observations of Geminga and Monogem. These observed TeV halos must instead occupy unusual regions, distinct from the majority of the local ISM.

\begin{figure}
\includegraphics[width=3.80in,angle=0]{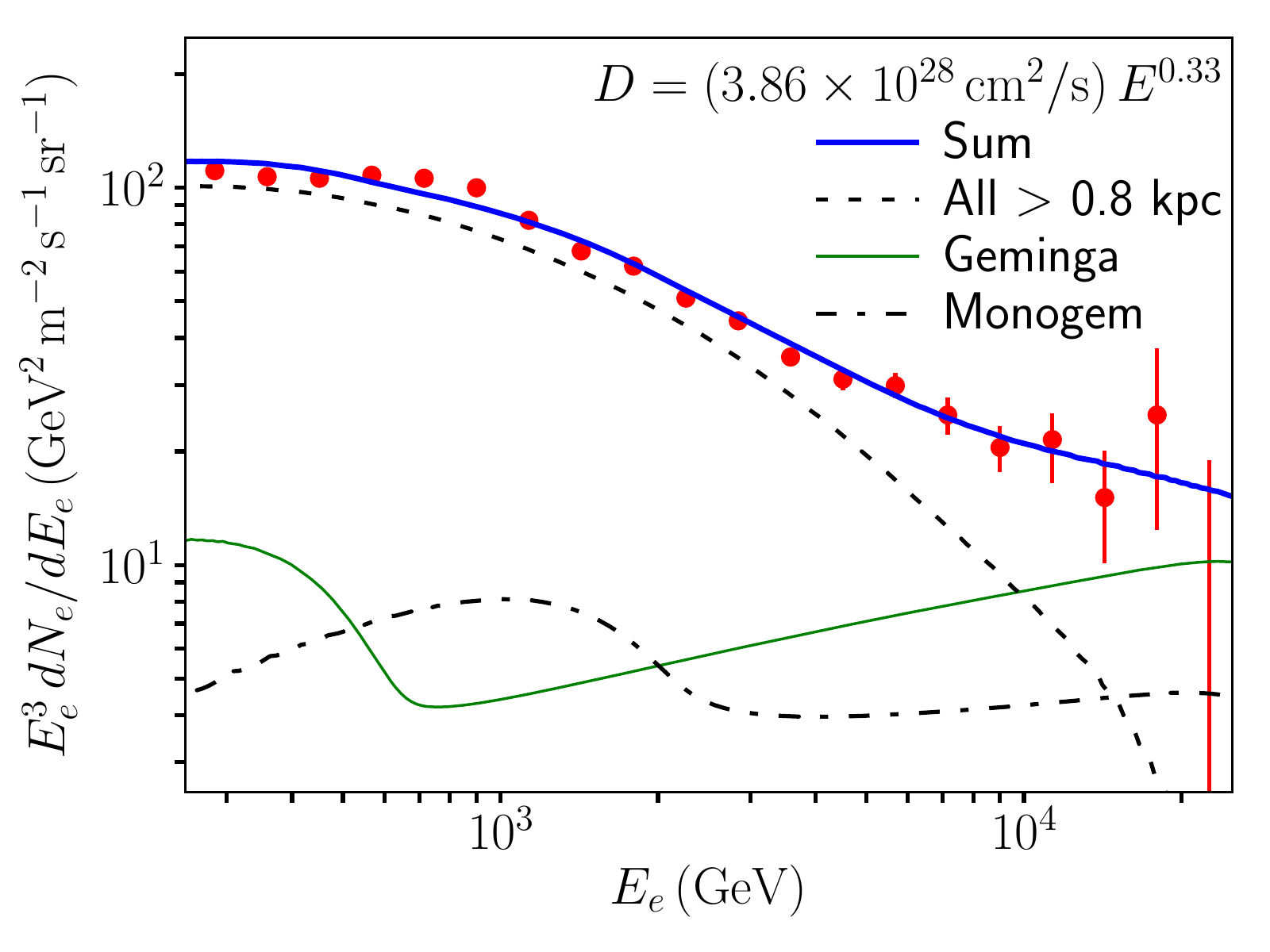}
\caption{The contribution to the cosmic-ray electron (plus positron) spectrum from all sources more than 800 pc from the Solar System (dashed black), and from the Geminga (dashed green) and Monogem (dot-dashed black) pulsars. Here we had adopted a distribution of sources with a scale height of 300 pc, injected spectral indices of $\alpha=2$, pulsar spindown timescales of $\tau=10^4$ years, and a diffusion coefficient of $D=3.86 \times 10^{28} \, (E_e/{\rm GeV})^{0.33}$ cm$^2/$s. This collection of sources can easily account for the cosmic-ray electron spectrum as measured by HESS and other experiments, while scenarios with a much lower diffusion coefficient cannot.}
\label{pop}
\end{figure}

Finally, we consider a simple but realistic distribution of cosmic-ray sources in order to demonstrate that the observed electron spectrum can be easily understood within the context of conventional diffusion models, such as those long-favored by the boron-to-carbon ratio and other measurements. We adopt a distribution of cosmic-ray sources that is described by a Lorimer profile~\cite{Lorimer:2003qc} with an exponential disk:
\begin{equation}
n_{\rm sources} \propto R^{2.35} \, \exp(-R/1530 \,{\rm pc}) \, \exp(-|z|/300 \,{\rm pc}).
\end{equation}
Here $R$ and $z$ describe the Galaxy in cylindrical coordinates. We also adopt an injected spectral index of $\alpha=2$ and a standard diffusion coefficient of \mbox{$D=3.86 \times 10^{28} \, (E_e/{\rm GeV})^{0.33}$ cm$^2/$s.} In Fig.~\ref{pop}, the black dashed line denotes the contribution from the portion of this source population that is located at more than 0.8 kpc from the Solar System, which we assume to be in steady-state. From this figure, it is clear that such a population can account for the observed cosmic-ray electron spectrum up to $\sim$3 TeV. At higher energies, however, local sources must play an important role. 

With this in mind, we additionally show the contributions to the cosmic-ray electron spectrum from the Geminga (dashed green) and Monogem (dot-dashed black) pulsars, adopting ages and distances for these sources of 370 and 110~kyr, and 250 and 280~pc, respectively. For these individual systems, we have assumed a time-dependent emission intensity that is proportional to the spin-down power of the pulsar, $L_e \propto [1+(t/\tau)]^{-2}$, where we have chosen a spin-down timescale of $\tau = 10^4$ years~\cite{Gaensler:2006ua}. In each case, we have normalized the emission to 30\% of the pulsars' total spindown power and have adopted an injected spectral index of $\alpha=2$. We note that the injected spectrum of $\alpha=2$ is somewhat softer than the 1.5~$<$~$\alpha$~$<$~1.9 spectrum necessary to fit the gamma-ray emission spectrum observed by HAWC from these sources~\citep{Hooper:2017gtd}. However, this softened spectrum approximately takes into account the additional electron cooling which occurs within the TeV halo before these electrons escape into the surrounding ISM. The sum of the contributions from these two sources and that from the population of more distant sources can easily account for the spectrum that has been reported by HESS and other experiments.  In contrast, if we adopt a diffusion coefficient that is $\sim$100-500 times lower (as argued in Ref.~\cite{newhawc}), it is not possible to account for the cosmic-ray electron spectrum observed above $\sim$1-2 TeV.
 
\section{Discussion and Summary}

Recent observations by the HAWC Collaboration of the very high-energy (VHE) gamma-ray emission from the Geminga and Monogem pulsars indicate that cosmic rays diffuse very slowly in the regions immediately surrounding these objects. The HAWC Collaboration has recently interpreted these measurements as evidence for a very low diffusion coefficient throughout the local interstellar medium (ISM)~\cite{newhawc}, approximately $\sim$100-500 times smaller than the value long-inferred from measurements of boron-to-carbon and other secondary-to-primary ratios in the cosmic-ray spectrum. If this were the case, the positron excess as reported by PAMELA~\cite{Adriani:2010rc} and AMS-02~\cite{Aguilar:2013qda} could not be accounted for by pulsars, motivating more exotic explanations.

If the value of the local diffusion coefficient were as low as advocated by the HAWC Collaboration, some exotic mechanism would be required in order to produce the observed positron excess. In such a scenario, we point out that this mechanism, or a second exotic mechanism, would also be required to produce the measured spectrum of VHE electrons. Such scenarios are very strongly constrained, however. For the case of annihilating dark matter, for example, gamma-ray observations of dwarf spheroidal galaxies by the Fermi-LAT Collaboration have already ruled out the vast majority of dark matter models that are potentially capable of generating the observed positron excess~\cite{Geringer-Sameth:2014qqa,Ackermann:2015zua,Fermi-LAT:2016uux}. Furthermore, the rapid cooling times of $\sim$20 TeV electrons require even more massive dark matter particles, with significantly higher annihilation rates. Gamma-ray constraints thus extremely strongly disfavor dark matter interpretations of the VHE cosmic-ray electron spectrum.

In this paper, we have argued that another interpretation of this data is much more likely. In particular, we have demonstrated that the cosmic-ray electron spectrum as measured by several experiments including HESS~\cite{hesstalk} is incompatible with scenarios in which transport through the ISM is described by a diffusion coefficient that is as low as suggested by the HAWC Collaboration~\cite{newhawc}. More specifically, with such inefficient diffusion, the highest energy electrons currently observed by HESS ($\sim$20 TeV) would only travel $\sim$10-20 pc before losing their energy. Given that no plausible sources of VHE cosmic rays exist within this radius, one must conclude that transport throughout the bulk of the ISM is described by a substantially larger diffusion coefficient.

HAWC's observations inform us as to the diffusion coefficient within approximately $\sim$25 pc around Geminga and Monogem, but provide no direct information about diffusion in the remaining bulk of the ISM. We argue here that it is very likely that the conditions that dictate cosmic-ray diffusion around Geminga and B0656+14 are very different from those found elsewhere in the ISM. Intriguingly, previous work has suggested that cosmic-ray gradients can produce regions of inhibited diffusion near supernova remnants on much smaller distance scales~\citep{1981A&A....98..195A, 2008AdSpR..42..486P, 2013ApJ...768...73M, DAngelo:2015cfw, 2016MNRAS.461.3552N, DAngelo:2017rou}, and it seems plausible that an analogous mechanism could lead to low diffusion coefficients in the regions surrounding middle-aged pulsars.

If all, or most, young and middle-aged pulsars are surrounded by a region with inefficient diffusion ($D \ll D_{\rm ISM}$), the volume of such regions must be fairly small in order to avoid conflicting with measurements of boron-to-carbon and other secondary-to-primary ratios. In particular, if such regions have a typical radius of $r_{\rm region}$, then they will collectively occupy roughly the following fraction of the volume of the Milky Way's disk:~\citep{Hooper:2017gtd}
\begin{eqnarray}
f &\sim& \frac{N_{\rm region} \times \frac{4\pi}{3} r^3_{\rm region}}{\pi R^2_{\rm MW} \times 2 z_{\rm MW}} \\
&\sim& 0.007 \times \bigg(\frac{r_{\rm region}}{30 \, {\rm pc}}\bigg)^3 \, \bigg(\frac{\dot{N}_{\rm SN}}{0.03 \, {\rm yr}^{-1}}\bigg) \, \bigg(\frac{\tau_{\rm region}}{10^6 \, {\rm yr}}\bigg) \, \bigg(\frac{20 \, {\rm kpc}}{R_{\rm MW}}\bigg)^2 \, \bigg(\frac{200 \, {\rm pc}}{z_{\rm MW}}\bigg), \nonumber
\end{eqnarray}
where $\dot{N}_{\rm SN}$ is the rate at which new pulsars appear in the Galaxy, $\tau_{\rm region}$ is the length of time that such regions persist, and $N_{\rm region} = \dot{N}_{\rm SN} \times \tau_{\rm region}$ is the number of such regions present at a given time. The quantities $R_{\rm MW}$ and $z_{\rm MW}$ denote the radius and half-width of the Galaxy's cylindrical disk. For such a small fraction of the total volume of the ISM, we expect such regions to have little impact on the observed secondary-to-primary ratios. In contrast, if most of such regions were as large as $r_{\rm region}\sim 150$ pc or greater, they would collectively occupy most of the ISM and dramatically impact cosmic-ray transport throughout the Milky Way. 



Lastly, we would like to emphasize that we are not arguing in this paper that the diffusion coefficient is entirely uniform or homogeneous throughout the Milky Way's ISM. Some variation is, of course, almost certain to exist. It has even been shown that the measured spectra of cosmic-ray protons, antiprotons and helium nuclei prefer somewhat different propagation parameters than those which provide the best-fit to the observed ratios of heavier nuclei (Be, B, C, N, O). This suggests that different cosmic-ray species may be probing different regions of the ISM, with different physical characteristics~\cite{Johannesson:2016rlh}. These differences, however, are much more modest than those being argued for in Ref.~\cite{newhawc}, being on the scale of tens of percents rather than factors of $\sim$100-500. The fact that the spectrum of each of these cosmic-ray species favors similar (if not identical) diffusion parameters to those shown here to be favored by the cosmic-ray electron spectrum provides considerable evidence for approximately homogeneous diffusion throughout the bulk of the Milky Way's ISM.

{\it In Summary}, the observations by HAWC and Milagro have revealed that the pulsars Geminga and Monogem are surrounded by regions in which diffusion is quite inefficient, featuring diffusion coefficients that are hundreds of times smaller than those determined by boron-to-carbon and other cosmic-ray secondary-to-primary ratios. Here we have argued, however, that these regions must be of quite limited size, and that diffusion must be significantly more effective throughout the majority of the local ISM. In particular, we show that the VHE electron spectrum can only be explained (without the presence of extremely nearby sources, $d \lsim 20$ pc) if the local diffusion coefficient is more similar to the value long-inferred from boron-to-carbon and other cosmic ray measurements than to that recently advocated for by the HAWC Collaboration.


\bigskip
\bigskip
\bigskip

\textbf{Acknowledgments.} DH is supported by the US Department of Energy under contract DE-FG02-13ER41958. Fermilab is operated by Fermi Research Alliance, LLC, under contract DE- AC02-07CH11359 with the US Department of Energy. TL acknowledges support from NSF Grant PHY-1404311.

\bibliography{hesselectron.bib}
\bibliographystyle{JHEP}

\end{document}